\documentstyle[11pt,amstex,righttag]{article}
\addtocounter{secnumdepth}{1}
\newcommand{\x}{\text{\bf x}}
\newcommand{\y}{\text{\bf y}}
\newcommand{\ba}{\text{\bf a}}
\newcommand{\bb}{\text{\bf b}}
\newcommand{\q}{\text{\bf q}}
\newcommand{\be}{\text{\bf e}}
\newcommand{\defin}{\equiv}
\newcommand{\bu}{\text{\bf u}}

\newcommand{\r}{\text{\bf r}}
\newcommand{\eps}{\varepsilon}
\numberwithin{equation}{section}
\begin{document}

\thispagestyle{empty}
\title{
Topology of the support\\ of the two-dimensional random walk
} 
\author{F. van Wijland,\, S. Caser,\, and H.J. Hilhorst\\
Laboratoire de Physique Th\'eorique et Hautes Energies\\
B\^atiment 211\\
Universit\'e de Paris-Sud\\
91405 Orsay cedex, France\\}

\maketitle
\vspace{-1cm}
\begin{small}
\begin{abstract}
\noindent We study the support ({\it i.e. the set of visited sites}) of a
$t$ step random walk on a two-dimensional square
lattice in the large $t$ limit. 
A broad class of global properties $M(t)$ of the support is considered,
including, {\it e.g.}, the number
$S(t)$ of its sites; the length of its boundary; the number of islands of
unvisited sites that it encloses; the number of such islands of given
shape, size, and orientation; and the number of occurrences in space of
specific local patterns of visited and unvisited sites.
On a finite lattice we determine the scaling functions that describe the 
averages $\overline{M}(t)$ on
appropriate lattice size dependent time scales. On an infinite
lattice we first observe that the  
$\overline{M}(t)$ all increase with $t$  as
$\sim t/\log^k t$, where $k$ is an $M$ dependent positive integer. We
then consider the class of random processes constituted by 
the {\it fluctuations around average} $\Delta M(t)$.
We show that to leading order as $t$ gets large these fluctuations are all
proportional to a {\it single universal random process} $\eta(t)$,
normalized to $\overline{\eta^2}(t)=1$.  
For $t\rightarrow\infty$ the probability law of $\eta(t)$ tends
to that of 
Varadhan's {\it renormalized local time of self-intersections}.
An implication is that in the long time limit all $\Delta M(t)$ are
proportional to $\Delta S(t)$. 

\end{abstract}
\end{small}
\vspace{3cm}
\noindent L.P.T.H.E. - ORSAY 96/72
\newpage
\section{Introduction and summary} 
\label{intoduction}
The reason for the multiple connections between the random walk and 
many questions of current and of permanent interest in science 
is evidently mathematical:
The random walk models the action of the Laplace operator, which is common to 
all those problems. Indeed, 
important early results on the random walk are due to
mathematicians. Among them, the famous Polya theorem \cite{Polya}
asserts that in spatial dimensions $d \leq 2$ a random walk is
certain to return to its initial position, whereas in $d>2$
it will escape to infinity. The random walk in $d=2$ is therefore at its 
critical dimension for return to the origin. Many features
associated with this fact 
make the random walk in two dimensions particularly 
interesting. For example, the probability distribution of the time
interval $\tau_0$ 
between two successive visits of the walk to its initial position decays
for large intervals as ~$\sim\! 1/\tau_0 \log^2 \tau_0$, so that very
long excursions occur away from the point of departure.
Excellent recent monographs on random walk theory have been written
by Weiss \cite{Weiss} and by Hughes \cite{Hughes}.\\

In this work we consider the {\it simple random walk} on a
two-dimensional square lattice: The
walk starts at time $t = 0$ at the origin {\bf x} $=$ {\bf 0} and steps at 
$t = 1, 2, 3,\ldots$ with equal probability to one of the 
four neighboring lattice sites. Our investigation focuses on 
the statistical properties of the {\it support}
of the walk at time $t$, {\it i.e.} of the set of sites that have
been visited during the first $t$ steps. 

At any given time $t$ the support can be
visualized as a set of black sites (the visited ones)  
in a lattice of otherwise white (unvisited) sites, as shown in Fig.\,1.
The set of unvisited sites is divided into components that are disjoint 
({\it i.e.} not connected via any nearest neighbor link) and that we  
call {\it islands}. 
In the course of time existing islands will be reduced in
size and single-site islands will eventually be destroyed; the number of
islands increases each time that 
a step of the walk cuts an existing island, or the outer region
surrounding the support, into disjoint components.

Questions about the statistical properties of the islands 
are natural and our interest in them 
was raised by a simulation study by Coutinho, Coutinho-Filho,
Gomes, and Nemirovsky \cite{Coutinho}. These authors 
investigated the evolution of the number of islands $I(t)$ and several
of their properties on {\it finite} lattices of up to $1200^2$ sites. 
In a short report \cite{CaserHilhorst} 
we showed that an analytic calculation is
possible, both on finite and infinite lattices, 
for {\it some} (not all) of the quantities considered by
Coutinho {\it et al.} \cite{Coutinho}, and that good agreement between 
theory and computer simulation is obtained. 
In this work we present a full account of most of the results announced in
Ref.\,\cite{CaserHilhorst} and consider a great many related questions.
We are, in particular, led to consider the infinite lattice again.\\

It appears that the number of islands $I(t)$ 
is but one member of a much
wider class of observables, generically to be denoted as $M(t)$, 
with closely related properties. 
This class includes the total number $S(t)$ of sites in the
support as well as the total length $E(t)$ of the boundary of the 
support (on an infinite lattice the boundary length is the sum of the external
perimeter and the perimeters of the islands enclosed).\\

The main variable that characterizes the support is its total number of
sites $S(t)$, sometimes called its {\it range}, which was first studied 
by Dvoretzky and Erd\"os \cite{DvoretzkyErdos}.
In the limit of large times $t$ it has the average value
\cite{DvoretzkyErdos,Vineyard,Montroll,
MontrollWeiss}
\begin{equation}
\overline{S}(t) \simeq \frac{\pi t}{\log 8t}
\label{Saverage}
\end{equation}
and the root mean square deviation \cite{JainPruitt, Torney}
\begin{equation}
\overline{\Delta S^2(t)}^{1/2}\simeq \cal{A}\frac{\pi t}{\log^2 8t}
\label{Svariance}
\end{equation}
where $\cal{A}=1.303...$ and where we write $\Delta S \equiv
S-\overline{S}$.  The typical support is known to be far from spherical, 
and its principal moments of
inertia and asphericity have been studied \cite{RudnickGaspari,Sciutto}.
Expressions exist \cite{WeissHavlin} also for its {\sl span}, that
is, the smallest rectangular box that it fits in.\\

The question of calculating the average $\overline{S}(t)$ 
becomes, in general dimension and in an appropriately taken continuum limit, 
the celebrated {\it Wiener sausage} problem: What is
the volume swept out in time $t$ by a $d$-dimensional Brownian sphere 
of finite radius? The Wiener sausage appears
naturally in certain applications of the random walk such as, for
example, the study by Kac and Luttinger \cite{KacLuttinger} of
Bose-Einstein condensation in the presence of impurities.
Historically this continuum problem precedes its
lattice counterpart. 
It was considered as early as 1933 by Leontovitsh
and Kolmogorov \cite{LeontovitshKolmogorov} and is still today an active 
subject of investigation in probability theory (\cite{LeGall}; see 
\cite{LeGallcourse} for a recent 
overwiew) and in mathematical physics \cite{Berezhkovskiietal}. 

The islands in the support of a lattice random
walk have their counterpart in the connected components into
which a two-dimensional Brownian motion path divides the plane. 
These have been studied by Mountford \cite{Mountford} and  Le Gall 
\cite{LeGall}, and, most recently, by Werner \cite{Werner}, who determines,
among other things, how many
there are larger than a given size $\epsilon$, in the limit of
$\epsilon\rightarrow 0$. The outer boundary of the Brownian motion path
has been considered very recently by Lawler \cite{Lawler}.\\    

We characterize as follows the class of observables $M(t)$ to be studied
below. For each lattice site $\x$ we introduce an 
{\it occupation number} $m(\x,t)$, equal to 0 if site {\bf x}  
is visited by the walker before or at time $t$, and equal to 1 if it
is not. 
Let now $A_1$ and $A_2$ be disjoint finite sets of lattice vectors.
Then the product 
\begin{equation}
\prod_{\ba_1\in A_1}m(\x+\ba_1,t)\prod_{\ba_2\in A_2}[1-m(\x+\ba_2,t)] 
\label{pattern}
\end{equation} 
codes for a specific spatial {\it pattern} $\alpha\equiv(A_1,A_2)$ of white 
(unvisited) and black (visited) sites.  
When $\x$ runs through the lattice, the product (\ref{pattern}) equals 1 
when the pattern $\alpha$ is encountered and
0 otherwise; hence this product summed on all $\x$ represents the 
total number $N_\alpha(t)$ of
occurrences in space of the pattern $\alpha$.
The observables that this work deals with are
these {\it pattern numbers} $N_\alpha(t)$ and their linear combinations $M(t)$.
It is easy to see that suitably chosen $M(t)$ may represent, {\it e.g.},
the total number 
$S(t)$ of sites in the support; the total boundary length $E(t)$ of the
support; the total number $I(t)$ of islands of unvisited sites enclosed by the
support; or the total number $I_\beta(t)$ of islands of a given type
$\beta$ (where {\it type} stands for shape, size,
and orientation).
Our main results concerning the observables $M(t)$ are of two kinds.

{\it (i) Average behavior on a finite lattice.}\;
On a finite lattice of $N$ sites the averages $\overline{M}(t)$ 
approach their limiting values on the time scale $t\sim N\log^2 N$,
where\-as their 
main vari\-ation oc\-curs on the ear\-lier time scale $t\sim N\log N$.  We 
calculate the scaling functions that describe the time dependence on 
both time scales. Several of the observables $M(t)$ of interest just
mentioned 
are treated as examples. A comparison is made, where possible, with the
simulations by Coutinho {\it et al.} \cite{Coutinho}

{\it (ii) Average behavior and fluctuations on the infinite
lattice.}\; 
On an infinite lattice the fluctuating
properties of the support manifest a universality that can be described
as follows. Let $M$ and $M'$ be 
two linear combinations of pattern numbers.
Then, with the same notation as before for deviations from average, 
we show by explicit calculation that asymptotically for $t\rightarrow\infty$
\begin{equation}
\overline{M}(t) \simeq m_k\frac{\pi^{k+1} t}{\log^{k+1} 8t}, \; \qquad
\overline{M'}(t) \simeq m'_{k'}\frac{\pi^{k'+1} t}{\log^{k'+1} 8t}
\label{Naverage}
\end{equation}
\begin{equation}
\overline{\Delta M(t)\Delta M'(t)} \simeq
4\cal{A}^2 (k+1)(k'+1) m_k m'_{k'} \frac{\pi^{k+k'+2} t^2}{\log^{k+k'+2} 8t} 
\label{Nvariance}
\end{equation}
Here $k$ and $k'$ are nonnegative integers that depend on $M$ and $M'$,
respectively, and are called their {\it order}\,; 
$m_k$ and $m'_{k'}$ are known proportionality constants; and
the same number $\cal{A}$ appears that was first
encountered in the study of the variance of $S(t)$.
The above equations imply that the normalized deviations from average
\begin{equation}
\eta_M(t) = \frac{\log 8t}{(k+1)\cal{A}}\frac{\Delta M(t)}{\overline{M}(t)}
\label{defetaMintro}
\end{equation}
have a correlation matrix $\overline{\eta_M(t)\eta_{M'}(t)}$ whose
elements all equal unity. This can be true only if all $\eta_M$
are equal to a {\it single random variable} to be called $\eta(t)$.
As a consequence the random variables $\Delta M(t)$ are, to leading order
as $t\rightarrow\infty$, all proportional to  
$\eta(t)$. Explicitly,
\begin{equation}
\Delta M(t)\simeq (k+1)m_k\frac{\pi^{k+1}t}{\log^{k+2} 8t}\,\cal{A}\,\eta(t)
\label{DMetaintro}
\end{equation}  
where $\eta(t)$ is {\it universal} \,(independent of $M$).
Hence what seemed to be a large number of independent fluctuating 
degrees of freedom of the support 
is hereby reduced, to leading order as $t\rightarrow\infty$, to only a
single degree of freedom.\\

We now briefly summarize the contents of the successive sections.
Since all quantities of interest are, in the end, expressed
in terms of the random walk Green function, we collect in 
Sec.\,\ref{Greenf} the basic 
knowledge required about this function. In
Sec.\,\ref{islandsfr} we express the total boundary length $E(t)$ and 
the total number
of islands $I(t)$ in terms of the occupation numbers 
and discuss the wider class of observables $M(t)$ of which they are examples.
In Secs.\,\ref{relfirst} and \ref{firstpassage} we calculate
the time dependent averages $\overline{M}(t)$ of such
observables. In
Secs.\,\ref{longinfinite}--\ref{examples2} the general expression for
the result, Eq.\,(\ref{resultMtaverage}), is analyzed 
for large times and explicitly worked out for 
several examples both on the infinite and the finite 
lattice. In Secs.\,\ref{relfirst2} and \ref{sclimitlongtime} we show, for the
infinite lattice, how to calculate the correlation between two observables
$M(t)$ and $M'(t)$. 
In Sec.\,\ref{conclusions} we arrive at the final results concerning the random
variable $\eta(t)$.
In the discussion in
Sec.\,\ref{discussion} we present various comments, compare where
possible our lattice results to  
their continuum analogs, and speculate about further connections between
several quantities.

\section{The random walk Green function}
\label{Greenf}
\subsection{Definition}
\label{basicdef}
A random walker starts at time $t=0$ at the origin $\x=\text{\bf 0}$ of a
square lattice and steps at each instant of time $t=1,2,3,\ldots$ with
probability $\frac{1}{4}$ to one of its four nearest-neighbor sites. The
Green function  
$G(\x,t)$ denotes the probability that at time $t$ the walker is at
lattice site $\x$, and  
\begin{equation}
\hat{G}(\x,z) = \sum_{t=0}^\infty z^t G(\x,t)
\label{definGxz}
\end{equation}
its generating function. In this section we 
collect in concise form those 
properties of the generating function that will be needed later.
An elementary calculation gives,
for a finite periodic lattice of $L\times L = N$ sites,
\begin{equation}
\hat{G}(\x,z) = \frac{1}{N} \sum_{\q} \frac{e^{-i\q\cdot \x}}{1
- \frac{1}{2}z(\cos q_1 + \cos q_2)}
\label{Gfinite}
\end{equation}
where $\q = (q_1,q_2) = 2\pi(\kappa_1,\kappa_2)/L$ with the $\kappa_i$
running through the values $0,1,2,\ldots,L-1$. In the limit of an
infinite lattice expression (\ref{Gfinite}) becomes
\begin{equation}
\hat{G}(\x,z) = \frac{1}{(2\pi)^2}\int_{-\pi}^\pi dq_1 \int_{-\pi}^\pi
dq_2\: \frac{e^{-i\q\cdot \x}}{1
- \frac{1}{2}z(\cos q_1 + \cos q_2)} 
\label{Ginfinite}
\end{equation}
At each point in this work it will be clear whether we are
discussing the finite or the infinite lattice; in some cases we shall denote 
corresponding quantities in the two geometries by the same symbol, as
for example 
in Eqs.\,(\ref{Gfinite}) and (\ref{Ginfinite}), and not explicitly
indicate their $N$ dependence on a finite lattice.

\subsection{Expansion near $z=1$}
\label{zto1}
The long time behavior of the physical quantities of interest is
determined by the behavior of $\hat{G}(\x,z)$ in the complex plane
near $z=1$. Expressions (\ref{Gfinite}) and (\ref{Ginfinite}) both have
the property that for $z\rightarrow 1$ the function $\hat{G}(\x,z)$
diverges. In order to study this divergence it is convenient to write
\begin{equation}
\hat{G}(\x,z)=\hat{G}(\text{\bf 0},z)-g(\x,z)
\label{GG-g}
\end{equation}
where on the RHS the term $g(\x,z)$ contains all the $\x$ dependence
and remains finite for $z\rightarrow 1$. We shall discuss
$\hat{G}(\text{\bf 0},z)$ and $g(\x,z)$ separately.
\subsubsection{The function $\hat{G}(\mbox{0},z)$} 
{\sl Finite lattice.} \, Expression (\ref{Gfinite}) has a simple pole
as a function of $z$ whenever one of the denominators inside the sum on
$\q$ vanishes. This leads to a sequence of poles on the real axis for
$z\geq 1$, of which the first one is located exactly at $z=1$. The
interval between two successive poles is of ${\cal O}(N^{-1})$ and
contains  
a zero. Upon setting $\x=\text{\bf 0}$ in Eq.\,(\ref{Gfinite}) and
expanding each term for small $1-z$ one gets \cite{denHollanderKasteleyn}
\begin{equation}
\hat{G}(\text{\bf 0},z)=\frac{1}{N(1-z)}+a(N)-a_1(N)(1-z)+{\cal O}\big((1-z)^2\big)
\label{Gfinite2}
\end{equation}
in which the coefficients are functions of $N$ that in the limit
$N\rightarrow \infty$ behave as \cite{denHollanderKasteleyn}
\begin{align}
a(N) = &\: \frac{1}{\pi}\log cN+{\cal O}(N^{-1})\nonumber\\
a_1(N) = &\: c_1 N+{\cal O}(\log N)  
\label{coefGfinite}
\end{align}
with $c=1.8456...$\, and $c_1=0.06187...$\\

The expansion in Eq.\,(\ref{Gfinite2}) represents well the behavior of
$\hat{G}(\text{\bf 0},z)$ near the pole at $z=1$, but is
certainly not valid on approach of the next pole. It can be used, however, to
determine the location of the zero $z=z_0$ between the first two poles.
Upon solving Eq.\,(\ref{Gfinite2}) 
in successive orders for $z_0$ one finds
\begin{equation}
z_0=1+\frac{1}{Na(N)}\Big[1-\frac{c_1}{a^2(N)}+\cdots \Big]
\label{z0}
\end{equation}  
From Eqs.\,(\ref{z0}) and (\ref{coefGfinite}) we conclude that
when $N\rightarrow \infty$ this zero is separated from the pole at
$z=1$ by a distance only of ${\cal O}(1/N\log N)$. 
All other zeros of $\hat{G}$ are separated from $z=1$ by a distance
of at least ${\cal O}(N^{-1})$. Due to its exceptional proximity to $z=1$
the zero $z_0$ plays a special role in the long-time behavior of the
random walk on finite lattices. This fact was first
noted by Weiss, Havlin, and Bunde \cite{WeissHavlinBunde} and has
also been exploited \cite{BrummelhuisHilhorst} in the study
of the covering time of a finite lattice by a
random walk.
\\
\noindent{\sl Infinite lattice.} \, When $N\rightarrow\infty$ the poles of
$\hat{G}$ densify to a branch cut and the expansion near $z=1$ is
\cite{ZumofenBlumen}
\begin{equation}
\hat{G}(\text{\bf 0},z) = \frac{1}{\pi} \log
\frac{8}{1-z} +{\cal O}\Big((1-z)\log(1-z)\Big)
\label{Ginfinite2}
\end{equation}
Later on in this work we shall also for brevity denote
$\hat{G}(\text{\bf 0},z)$ as $G_0(z)$.

\subsubsection{The function $g(\mbox{x},z)$}

Expanding $g(\x,z)$ for finite $N$ around $z=1$ gives
\begin{equation}
g(\x,z)= g_N(\x)+g_N'(\x)(1-z)+{\cal O}((1-z)^2)
\label{gfinite2}
\end{equation}
where we now explicitly indicate that the expansion coefficients are
$N$ dependent. We shall set
\begin{equation}
g(\x)=\lim_{N\rightarrow\infty}g_N(\x).
\label{g}
\end{equation} 
Spitzer \cite{Spitzer} shows how to calculate the $g(\x)$ for $\x$
close to the origin. Letting $\text{\bf
e}_1$ and $\text{\bf e}_2$ denote the unit vectors we have the following
values: 
\begin{equation}
\begin{array}{lll}
g(\text{\bf 0})=0       &g(2\be_1)=4-8/\pi \\
g(\be_1)=1       &g(2\be_1+\be_2)=8/\pi-1 \\
g(\be_1+\be_2)=4/\pi   &
\end{array}
\label{gvalues}
\end{equation}
When combining preceding results we see that for finite $N$ the quantity
$\hat{G}(\x,z)$ has the expansion
\begin{equation}
\hat{G}(\x,z) \simeq \frac{1}{N(1-z)} + [a(N)-g_N(\x)] + {\cal O}(1-z)
\label{G2}
\end{equation}
whose second term behaves for large $N$ as
\begin{equation}
a(N)-g_N(\x) \simeq \frac{1}{\pi}\log cN - g(\x) +\cdots
\label{coeffGxfinite}
\end{equation}
with the dots representing terms that vanish as $N\rightarrow\infty.$

\subsection{Scaling limit}
\label{sclimit}

In our study of the fluctuations in Sec.\,\ref{fluctuations}
we shall also 
need $\hat{G}(\x,z)$ in the scaling limit
$z\rightarrow1, \; x\rightarrow\infty$ with $x^2(1-z)$ fixed. The behavior
in this limit is \cite{MontrollWeiss}
\begin{equation}
\hat{G}(\x,z) \simeq \frac{2}{\pi}K_0(2x(1-z)^{1/2})
\label{Gscaling}
\end{equation} 
where $K_0$ is the modified Bessel function of order zero. As is
well-known, for $z\rightarrow 1$ the dominant contribution to the sum
(\ref{definGxz})  comes from values of $t$ that are of ${\cal O}((1-z)^{-1})$,
and therefore the scaling limit corresponds to focusing on distances $x$
of ${\cal O}(\sqrt{t})$.
\section{Islands and other observables}
\label{islandsfr}

\subsection{Islands}
\label{islands}

The representation of Fig.\,1 divides the lattice into a black area and white
ones. We introduce for each lattice site {\bf x} the {\it occupation
number} 
\begin{equation}
m(\x,t)=\left\{ \begin{array}
{cl}1 & \text{if {\bf x} has not yet been visited at time } t
\\ 0 & \text{otherwise}
\end{array} \right.
\label{definm}
\end{equation}
At this point it may seem more natural to work with the
equivalent occupation numbers $n(\x,t)\defin 1-m(\x,t)$, but 
definition (\ref{definm}) will soon prove to be more convenient.\\

If starting at an arbitrary element, one
follows the boundary such that 
the white sites are on the left and the black ones on the right, then
one will return to the point of departure after having turned either through an
angle $2\pi$ (if that part of the boundary encloses an island in the
support) or through an angle $-2\pi$ (if it is the outer boundary of
the support). The number of islands is therefore obtained from
the number of turns in the boundary, by adding those of Fig.\,2a
with weight $+\frac{1}{4}$ and those of Fig.\,2b with weight $-\frac{1}{4}$.
The diagrams of Fig.\,2c correspond to two turns of $\pi/2$ and therefore
count with weight factor $\frac{1}{2}$. This procedure 
counts the outer boundary with weight $-1$ and therefore 2 has to be
added to obtain the final result.  
It is now easy to construct the expression 
for the total number of islands in terms of the occupation variables
$m(\x,t)$. We have to shift a $2\times 2$ window across the lattice, 
check all $2\times 2$ local site configurations and add all those that
are of the types of Fig.\,2 with the proper weights. Let $\x$
denote the lower lefthand site in the $2\times 2$ window. Then, for
example, the expression
\begin{equation}
m(\x,t)\,[1-m(\x+\be_1,t)]\,[1-m(\x+\be_2,t)]\,m(\x+\be_1+\be_2,t)
\label{examplem}
\end{equation}
(which is of type (\ref{pattern})) equals 1 or 0 according to whether
the local $2\times2$ configuration is
or is not equal to the diagram of Fig.\,2c. Writing down analogous
expressions for all other diagrams, summing these, and subsequently
summing them on $\x$ yields $I(t)$ as a linear combination of pattern numbers.
After rearranging terms one finds
\begin{equation}
\begin{split}
I(t) =\: & 2 + \sum_{\x} \Big[m(\x,t) - m(\x,t)m(\x+\be_1,t) -
m(\x,t)m(\x+\be_2,t) \\&+
m(\x,t)m(\x+\be_1,t)m(\x+\be_2,t)m(\x+\be_1+\be_2,t)\Big]
\end{split}
\label{exprI}
\end{equation}
This expression is at the basis of all calculations concerning islands on
the infinite lattice. Finite lattice calculations require a further
remark, which is made below Eq.\,(\ref{Imainlong}).

\subsection{Other observables}
\label{otherobservables}

We are now interested in other variables whose
values can be obtained by the "window" method. To see
the general form of these variables, let
$A$ be a finite subset of lattice vectors. The set $\{\x+\ba\,|\,
\ba \in A\}$, obtained by translating $A$ by a vector
$\x$, will be written as $\x+A$. The variable
\begin{equation}
m_{\x+A}(t) =  \prod_{\ba\in A} m(\x+\ba,t)
\label{definmxA}
\end{equation}
is equal to unity if at time $t$ all the sites of this set are
white (unvisited), and is zero otherwise. The sum variable
\begin{equation}\label{MA(t)}
M_A(t) = \sum_{\x} m_{\x+A}(t)
\label{definMA}
\end{equation}
counts the total number of wholly white sets in the
lattice that can be obtained from $A$ by a translation. 

In the remainder we shall also wish to take for $A$ the empty set
$\emptyset$. In that case the right hand side of Eq.\,(\ref{definmxA})
should be assigned the value unity and Eq.\,(\ref{definMA}) shows that
$M_{\emptyset}(t)$ is the total number of lattice sites.

The number of islands $I(t)$ was initially found in Sec.\,\ref{islands} as
a linear combination of pattern numbers, which led to the final
expression (\ref{exprI}). The representation of an observable as a 
linear combination of pattern numbers is in general nonunique, but
expression (\ref{exprI}) is a unique member of the 
class of variables $M(t)$ that are of the form
\begin{equation}\label{Generalform}
M(t) = \sum_A \mu_A M_A(t)
\label{definM}
\end{equation}
with arbitrary coefficients $\mu_A$. Various other
quantities of potential interest can be expressed this way. Some of
these, with their coefficients $\mu_A$, have been listed in Table I.
The best
known example is the total number $S(t)$ of sites in the support, 
\begin{equation}
S(t) = \sum_{\x}\, [ 1 - m(\x,t) ]
\label{exprSm}
\end{equation}
which has $\mu_A = \pm 1$ for $A = \emptyset$ and $A =\{\text{\bf
0}\}$, respectively,
and $\mu_A = 0$ otherwise.
Another example is the total boundary length $E(t)$ between the visited
and unvisited lattice sites (that is, the total number of pairs of
neighboring sites of which one is white and one black).  
It can be expressed as 
\begin{equation}
\begin{split}
E(t) & = \sum_{\x} \Big[ m(\x,t)(1-m(\x+\text{\bf e}_1,t)
+(1-m(\x,t))m(\x+\text{\bf e}_1,t)                                     \\
 & \quad +m(\x,t)(1-m(\x+\text{\bf e}_2,t)
+(1-m(\x,t))m(\x+\text{\bf e}_2,t) \Big]                                   \\
 & = \sum_{\x} \Big[ 4m(\x,t) - 2m(\x,t)m(\x+\text{\bf e}_1,t) -
2m(\x,t)m(\x+\be_2,t) \Big] \\
\label{definE}
\end{split}
\end{equation}
We note that one has the relation

\begin{equation}
\sum_A \mu_A = 0
\label{summuzero}
\end{equation}
for the three observables $S$, $E$, and $I$, but that
\begin{equation}
\sum_{A\neq\emptyset} \mu_A = 0
\label{sumneqmuzero}
\end{equation}
only for $E$ and $I$, but not for $S$. Property (\ref{summuzero}) is
required if the sum on $\x$ in Eq.\,(\ref{definMA}) is to have a finite
limit when the lattice size $N$ tends to infinity. The property (\ref{sumneqmuzero})
is easily traced back to the fact that for $E$ and $I$ the
window selects patterns consisting of both white and black sites,
whereas for $S$ it selects only black sites. When Eq.\,(\ref{summuzero}) holds,
Eq.\,(\ref{sumneqmuzero}) is equivalent to $\mu_\emptyset=0$.


\section{Averages of observables}
\label{calcaverages}
\subsection{Relation to first passage time probabilities}
\label{relfirst}

We shall now calculate in a unified way averages
and fluctuations of quantities $M(t)$ of type (\ref{definM}). 
The approach of this subsection will serve as
the basis for the developments to follow. Using
Eqs.\,(\ref{definmxA})--(\ref{definM}) we see that the averages
$\overline{M}(t)$   
are linear combinations of expressions of the type
\begin{eqnarray}\label{mx+A}
\overline{m_{\x+A}(t)} & = & \overline{\prod_{\ba\in A}
m(\x+\ba,t)}                                         \nonumber\\ 
& = & 1-\sum_{\tau=0}^t f_{\x+A}(\tau)          \nonumber\\
& = & 1-\sum_{\tau=0}^t \sum_{\ba \in A}
f_{\x+A}(\x+\ba,\tau)                                   
\label{mxAaverage}
\end{eqnarray}
in which the last two transformations make sense only when
$A\neq\emptyset$;
$f_{\x+A}(\tau)$ is the probability that the walker's first visit
to any site of the set $\x+A$ takes place at time
$\tau$; and $f_{\x+A}(\x+\ba,\tau)$ is the probability that it
takes place at time $\tau$ {\it and} that it concerns
the specific site $\x+\ba$. Upon averaging Eq.\,(\ref{definMA}), using
Eq.\,(\ref{mxAaverage}), and passing to generating functions we
find, for all nonempty $A$, 
\begin{equation}
\hat{\overline{M}_A}(z) = -\frac{1}{1-z}\sum_{\x}
\Big[\sum_{\ba\in A}\hat{f}_{\x+A}(\x+\ba,z) - 1\Big]
\label{MAzaverage}
\end{equation}
For the empty set one derives directly that
\begin{equation}
\hat{\overline{M}}_{\emptyset}(z)=N/(1-z)
\label{exprMempty}
\end{equation}
We now sum the $\hat{\overline{M}}_A(z)$ given by
Eqs.\,(\ref{MAzaverage}) and (\ref{exprMempty}) on all $A$ with
coefficients $\mu_A$. Using Eq.\,(\ref{summuzero}) 
and introducing for all $A\neq\emptyset$
\begin{equation}
\hat{F}_A(\ba,z) = \sum_{\x}\hat{f}_{\x+A}(\x+\ba,z)
\label{definFA}
\end{equation}
yields
\begin{equation}
\hat{\overline{M}}(z)=-\frac{1}{1-z}\sum_{A\neq\emptyset}\mu_A\sum_{\ba\in
A}\hat{F}_A(\ba,z) 
\label{exprMbarz}
\end{equation}
With this formula we have reduced the generating function of the average
of interest, $\overline{M}(t)$, to the quantities $\hat{F}_A$ which are closely
related to first passage times, but still unknown. We shall now 
proceed to determine the $\hat{F}_A.$

\subsection{Solving the first passage time probabilities}
\label{firstpassage}

Standard random walk theory \cite{Spitzer,Weiss} 
relates the first passage probabilities
$\hat{f}$ to the Green function $\hat{G}$ by

\begin{equation}
\hat{G}(\x + \ba,z) = \sum_{\ba'\in A}
\hat{f}_{\x+A}(\x+\ba')\hat{G}(\ba-\ba',z)
\label{eqnforf}
\end{equation}
for all $\ba\in A$. With the aid of Eq.\,(\ref{definFA}) we find for
$\hat{F}_A$ the equation 
\begin{equation}
\sum_{\ba' \in A} \hat{F}_A(\ba',z)\hat{G}(\ba-\ba',z) = \frac{1}{1-z}
\label{eqnforF}
\end{equation}
for all $\ba\in A$.
This is a matrix equation for the $\hat{F}_A$ whose dimension is the 
number $|A|$
of sites in the set $A$. This equation possesses special properties
which are best exhibited by converting it to the shorthand notation
\begin{eqnarray}
\gamma_{\ba\ba'}(z) & = & g(\ba-\ba',z)/G_0(z) \nonumber\\
{\cal F}_{\ba} & = & (1-z)G_0(z)\hat{F}_A(\ba,z)
\label{matrixshorthand}
\end{eqnarray}
Eq.\,(\ref{eqnforF}) then becomes

\begin{equation}
\sum_{\ba'\in A}(1-\gamma_{\ba\ba'}(z)){\cal F}_{\ba'} = 1
\label{eqnforcalF}
\end{equation}
for all $\ba\in A$. If we denote by $\gamma^{(A)}$ the matrix of
elements $\gamma_{\ba,\ba'}$ with $\ba,\,\ba'\in A$, then in
matrix notation
\begin{equation}
(J-\gamma^{(A)}(z)){\cal F} = \text{\bf j}
\label{matrixeqnforcalF}
\end{equation}
where $J$ and {\bf j} are the matrix and vector, respectively, of
dimension $|A|$, whose elements all equal 1. As shown in
Eq.\,(\ref{exprMbarz}), we only need the sum of the components of ${\cal F}$.
Formal inversion gives
\begin{equation}
\sum_{\ba\in A}{\cal F}_{\ba} = \sum_{\ba\in A}\sum_{\ba'\in A}
[(J-\gamma^{(A)}(z))^{-1}]_{\ba\ba'}
\label{solnsumcalF}
\end{equation}
In a later stage we shall wish to take the limit $z\rightarrow 1$.
In view of Eq.\,(\ref{matrixshorthand}) and the known behavior of
$G_0(z)$ this implies that $\gamma^{(A)}(z)\rightarrow 0$, 
so that, except when $|A|=1$, the matrix inverse $(J-\gamma^{(A)}(z))^{-1}$ in
Eq.\,(\ref{solnsumcalF}) 
ceases to exist.
We therefore now convert that equation to a form more suitable for
taking that limit. In the appendix it is shown that
\begin{equation}
\sum_{\ba\in A}{\cal F}_{\ba} = \frac{1}{1-\gamma_A}
\label{trfsumcalF}
\end{equation}
where
\begin{equation}
\gamma_A^{-1}(z) \defin \sum_{\ba\in A}\sum_{\ba'\in A}
[(\gamma^{(A)}(z))^{-1}]_{\ba\ba'}
\label{defingammaA}
\end{equation}
Upon coming back to the original notation, but now with the
abbreviation
\begin{equation}
g_A(z) \defin \gamma_A(z)G_0(z) \label{defingAG0}
\end{equation}
we find from Eq.\,(\ref{trfsumcalF}) the solution of Eq.\,(\ref{eqnforF}) in
the form
\begin{equation}
\sum_{\ba\in A}\hat{F}_A(\ba,z) =
\frac{1}{(1-z)[G_0(z)-g_A(z)]}
\label{solnsumF}
\end{equation}
When $|A|=1$, in which case we may take $A=\{{\text{\bf 0}}\}$, 
one deduces directly
from (\ref{eqnforF}) that (\ref{solnsumF}) holds with 
$g_{\{{\text{\bf 0}}\}}(z)=0$. 
For $A$ of diameter not too large, as is the case in
many examples of interest, the
quantity $g_A(z)$ is easily expressed explicitly in terms of the
$g(\ba-\ba',z)$. 
If after substitution of expression (\ref{solnsumF}) in Eq.\,(\ref{exprMbarz})
we sum on $A$, use Eq.\,(\ref{summuzero}), and transform back to the
time domain, we get
\begin{equation}
\overline{M}(t) = -\frac{1}{2\pi i}\oint
\frac{dz}{z^{t+1}}\frac{1}{(1-z)^2}
 \sum_{A\neq\emptyset} \mu_A \frac{1}{G_0(z)-g_A(z)}
\label{resultMtaverage}
\end{equation}
where the integral runs counterclockwise around the origin.
This result is still fully exact and applies to both finite and infinite
lattices.  

\subsection{Long-time behavior of averages. Infinite lattice}
\label{longinfinite}
The starting point for the analysis of this section is
Eq.\,(\ref{resultMtaverage}). 
On an infinite lattice the asymptotic behavior of $\overline{M}(t)$ as
$t\rightarrow\infty$ is determined
by the $z\rightarrow 1$ behavior of the integrand.
This behavior follows from expression (2.8) for
$G_0(z)$ and from Eqs.\,(\ref{defingAG0}) and
(\ref{defingammaA}) which together
determine $g_A(z)$. We shall satisfy ourselves with retaining 
the leading $z\rightarrow 1$ behavior {\it and} corrections
that are of relative order of negative powers of ~$\log(1-z)$.
The terms neglected are of relative order $1-z$, apart from logarithmic
factors. 
This means that in Eq.\,(\ref{resultMtaverage}) we may replace 
$G_0(z)$ with $\pi^{-1}\log(8/(1-z))$ and
the $g_A(z)$ with their values at $z=1$, which for brevity we shall
denote by $g_A$. The coefficient $g_A$ appears in potential theory and
is the two-dimensional lattice analog of the elec\-tro\-static capacity
of the set $A$; its properties have been
reviewed by Spitzer \cite{Spitzer}.

We expand the summand in
Eq.\,(\ref{resultMtaverage}) in inverse powers of $\log(8/(1-z))$. 
Using the above results we so obtain, asymptotically for 
$t \rightarrow \infty$, 
\begin{equation}
\overline{M}(t) \simeq \sum_{n=0}^{\infty} m_n {\cal J}_{n+1,2}(t)
\label{asptMJ}
\end{equation}
where the coefficients $m_n$ are determined by the observable $M$
according to
\begin{equation}
m_n = -\sum_{A\neq\emptyset}\mu_A g_A^n
\label{defcn}
\end{equation}
for $n=0,1,2,\ldots$ and where 
\begin{equation}
{\cal J}_{n\ell}(t) = \frac{1}{2\pi i} \oint \frac{dz}{z^{t+1}}
\frac{1}{(1-z)^{\ell}} \frac{\pi^n}{\log^n\!\frac{8}{1-z}} 
\label{defJnl}
\end{equation}
The $t\rightarrow \infty$ behavior of the integrals (\ref{defJnl})  
has been studied, in
particular, by Henyey and Seshadri \cite{HenyeySeshadri} for the case
$\ell=2$. A generalization of their result is 
\begin{equation}
{\cal J}_{n\ell}(t) \simeq \frac{\pi^n t^{\ell-1}}{\log^n 8t}
\sum_{m=0}^{\infty} (-1)^m 
\binom{m+n-1}{m}
\frac{b_{m\ell}}{\log^m 8t}
\label{asptJnl}
\end{equation}
with coefficients
\begin{equation}
b_{m\ell} = \frac{d^m}{dx^m} \frac{1}{\Gamma(x+\ell)}\Big|_{x=0}
\label{defbml}
\end{equation}
We have $b_{0\ell}=1/(\ell-1)!$ and shall also need explicitly below
\begin{equation}
\begin{array}{lll}
b_{12}&=-1+C&=-0.422...\\
b_{13}&=-\tfrac{3}{4}+\tfrac{1}{2}C&=-0.461...\\
b_{22}&=2-\tfrac{1}{6}\pi^2-2C+C^2&=-0.466...\\
b_{23}&=\tfrac{7}{4}-\tfrac{1}{12}\pi^2-\tfrac{3}{2}C+\tfrac{1}{2}C^2&
=\phantom{-}0.227...
\label{valuesbml}
\end{array}
\end{equation}
where $C=0.577215...$\, denotes Euler's constant.
Combining Eqs.\,(\ref{asptMJ}) and (\ref{asptJnl}) 
we find for $\overline{M}(t)$ an asymptotic expansion in inverse powers
of $\log t$, 
\begin{align}
\overline{M}(t)\simeq &\: \frac{\pi t}{\log 8t}\,m_0\nonumber\\
&+\frac{\pi t}{\log^2 8t}\,[-b_{12}m_0+\pi m_1]\nonumber\\
&+\frac{\pi t}{\log^3 8t}\,[b_{22}m_0-2\pi b_{12}m_1+\pi^2 m_2]\nonumber\\
&+\frac{\pi t}{\log^4 8t}\,[-b_{32}m_0+3\pi b_{22}m_1 - 3\pi^2 b_{12}m_2
+\pi^3 m_3]\nonumber\\
&+\ldots
\label{Mtfinal}
\end{align}
where we have set $b_{02}=1$.
We recall that whereas the $b_{m\ell}$ are numerical coefficients,
the $m_n$ defined by Eq.\,(\ref{defcn}) are specific for the
observable $M$.
Eq.\,(\ref{Mtfinal}) shows that the leading asymptotic behavior is $\,\sim
t/\log8t$ for observables that have $m_0\neq 0$, and $\,\sim t/\log^2
8t$ for those that have $m_0=0$ but $m_1\neq0$. 
In view of the definition of $m_0$ and the discussion at the end of
Sec.\,\ref{otherobservables},  
observables with $m_0\neq 0$ involve only black patterns and will be
called, for short, "black" observables, whereas those with $m_0=0$ 
will be called "black-and-white" observables.

\subsection{Examples}
\label{examples}
It is now easy to derive results for many examples of interest by
applying the formulas of Sec.\,\ref{longinfinite}.\\

\noindent{\it Example 1.1.}\; We take for $M$ the total
number $S$ of sites in the support. The first column of
coefficients $\mu_A$ in Table I
shows that the only nonzero term in the sum (\ref{defcn}) is due to the set
$A=\{\text{\bf 0}\}$. Since this set has $\mu_A=-1$ and $g_A=0$, the only
nonzero coefficient produced by Eq.\,(\ref{defcn}) is $m_0=1$. From
Eq.\,(\ref{Mtfinal}) we then have 
\begin{equation}
\overline{S}(t)\simeq\frac{\pi t}{\log 8t}\Big[1-\frac{b_{12}}{\log
8t}+\frac{b_{22}}{\log^2 8t}-\frac{b_{32}}{\log^3 8t}+\cdots\Big]
\label{Stfinal}
\end{equation}
The numerical values of the coefficients of the first two subleading
terms are given in Eq.\,(\ref{valuesbml}) and    
agree with those of Torney \cite{Torney}.\\ 

\noindent{\it Example 1.2.} \;Next we take for $M$ the total boundary
length $E$ between the 
white and black areas. In the second column of coefficients $\mu_A$ in
Table I shows that only two
sets $A$ enter, with the pair $(\mu_A,g_A)$ equal to $(4,0)$ and
$(-4,\frac 12)$. Eq.\,(\ref{defcn}) then leads to $m_0=0$ and $m_n=2^{2-n}$
for $n=1,2,\ldots$, after which 
Eq.\,(\ref{Mtfinal}) gives 
\begin{equation}
\overline{E}(t)\simeq
\frac{2\pi^2 t}{\log^2
8t}\Big[1+\frac{\eps_1}{\log 8t}+\frac{\eps_2}{\log ^2 8t}+\cdots\Big]
\label{Etfinal}
\end{equation}
with the coefficients 
\begin{eqnarray}
\eps_1&=&\tfrac{1}{2}\pi-2b_{12}=2.41636...\nonumber\\
\eps_2&=&\tfrac{1}{4}\pi^2-\tfrac{3}{2}\pi b_{12}+3 b_{22}=3.06116...
\label{Etcoeff}
\end{eqnarray}

\noindent{\it Example 1.3.}\; Now take for $M$ the total number $I$ of
islands. Using the third column of coefficients $\mu_A$ in Table I as input in
Eq.\,(\ref{defcn}) we find that $m_0=0$ and
\begin{equation}
m_n=2^{1-n}-[(\pi+2)/2\pi]^n
\label{Icn}
\end{equation}
for $n=1,2,\ldots$. Substituting as in the previous examples we find
\begin{equation}
\overline{I}(t)\simeq\tfrac{1}{2}\pi(\pi-2)\frac{t}{\log^2
8t}\Big[1+\frac{\iota_1}{\log 8t}+\frac{\iota_2}{\log^2 8t}+\cdots\Big]
\label{Itfinal}
\end{equation}
Analytic expressions for the coefficients $\iota_i$ are
easily found with the aid of the preceding formulas but are of little
interest. The numerical values of the first two of them are
\begin{eqnarray}
\iota_1&=& -\phantom{1}2.087...\nonumber\\
\iota_2&=& -21.304...
\label{Itcoeff}
\end{eqnarray}
In the leading asymptotic behavior of both $\overline{E}(t)$ and
$\overline{I}(t)$ 
an extra factor $1/\log 8t$ appears compared to that of $\overline{S}(t)$
as a consequence of $m_0$ being zero, {\it i.e.}, of $E(t)$ and $I(t)$ 
being black-and-white observables.
In Sec.\,\ref{longfinite} we shall come back to $\overline{I}(t)$ and
also make a 
comparison with the numerical simulations by Coutinho
{\it et al.} \cite{Coutinho}.\\

\noindent{\it Example 1.4.}\; Let $\beta$ be a specific type of island,
where {\it type} indicates shape, size and orientation. 
Let $I_\beta(t)$ be the observable that counts the total number of
islands  of that type. According to the preceding
discussion we must have that 
\begin{equation}
\overline{I}_\beta(t)\simeq \tfrac{1}{2}\pi(\pi-2)f_\beta\frac{t}{\log^2
8t}\;\;\;\text{ as }t\rightarrow \infty
\label{Ialphat}
\end{equation}
for some proportionality constant $f_{\beta}$,
even though we cannot expect the approach to this asymptotic  behavior
to be uniform in $\beta$. By summing Eq.\,(\ref{Ialphat}) on all
$\beta$ and comparing to Eq.\,(\ref{Itfinal}) we conclude that the
average number of islands of type $\beta$ represents, as
$t\rightarrow\infty$, a fixed {\it fraction} $f_\beta$ of the average total
number of islands.
We have calculated the fraction $f_1$ of islands that are 
single isolated sites and the fraction $f_2$ of {\it "dimer"} islands,  
consisting of two neighboring sites, with the result
\begin{align}
f_1&=-\frac{\pi(\pi^3-7\pi^2+14\pi-4)}{(\pi-1)(\pi^2-6\pi+4)}
=0.560079...\nonumber\\
f_2&=0.073557...    
\label{f1f2}
\end{align}
The main effort goes into the calculation of the necessary coefficients
$g_A$. Those needed for $I_1$ are listed in Table I. We do not present
the 22 coefficients needed for $I_2$, nor the final analytic expression
for $f_2$. Since the dimers may have two orientations, the islands of
sizes 1 and 2 represent a fraction $f_1+2f_2=0.707193...$ of all islands.\\

With these results in hand we return to the total boundary length
considered in Example 1.2.
On an infinite lattice one can
write $E(t)=E_{ext}(t)+E_{int}(t)$, where $E_{ext}(t)$ is the external
perimeter of the support and $E_{int}(t)$ the total perimeter of the
islands enclosed by it.
In Example 1.2. we determined only the average of their sum;
determining 
$\overline{E}_{ext}(t)$ and $\overline{E}_{int}(t)$ separately is
a much more difficult problem that we have not seen how to
solve by the present method.
A rigorous lower bound for $\overline{E}_{int}(t)$ is nevertheless easily
obtained by adding up the perimeters of the single-site and dimer
islands, and taking into account that all other islands have a perimeter
at least equal to 8. This gives $\overline{E}_{int}(t) >
0.4965...\overline{E}(t)$. By arguments different from ours Lawler
\cite{Lawler} shows that in fact for $t\rightarrow\infty$ the external
perimeter increases only as $\overline{E}_{ext}(t)\sim t^{2/3}$, so that
$\overline{E}_{int}(t)\simeq\overline{E}(t)$.\\ 

\noindent {\it Example 1.5.}\; Let finally $M(t)$ be equal to the
pattern number $N_\alpha(t)$ obtained by summing the expression
(\ref{pattern}) over all sites $\x$. If neither $A_1$ nor $A_2$ is
empty, then the pattern $\alpha$ is composed of both white and black
sites and has $m_0=0$. Hence the average total number of these patterns
is obtained by setting $m_0=0$ in Eq.\,(\ref{Mtfinal}) and increases
with $t$ as 
\begin{equation}
\overline{N}_\alpha(t) \simeq m_1 \frac{\pi^2 t}{\log^2 8t}
\label{avNalpha}
\end{equation}
where $m_1$ is $\alpha$ dependent. The black-and-white patterns
considered here necessarily lie on the boundary of the
support; comparison of Eqs.\,(\ref{avNalpha}) and (\ref{Etfinal}) shows
that the {\it number per unit of boundary length}
of such patterns tends to a fixed value as $t\rightarrow\infty$.

\subsection{Long-time behavior of averages. Finite lattice}
\label{longfinite}
For a finite lattice of $N$ sites, after an initial increase with time
identical to what happens on the infinite lattice,
we must expect deviations from the infinite lattice behavior to appear
on a characteristic $N$ dependent time scale
$\tau(N)$ that will tend to infinity when $N$ does.
For larger times all black observables  
will level off and 
tend to a constant times $N$, and all black-and-white 
observables will pass through a maximum value, then 
bend down and asymptotically approach zero. 
The expansion that we shall look for in the case of a finite
lattice will therefore involve a determination of this time scale.   

We shall not attempt in this case a full asymptotic expansion as 
for the infinite lattice, but only determine the leading asymptotic
behavior.
Our starting point is again Eq.\,(\ref{resultMtaverage}), in which, when
$z\rightarrow 1$, using
Eqs.\,(\ref{Gfinite2}) and (\ref{coefGfinite}), we 
may substitute
\begin{equation}
G_0(z)-g_A(z) = \frac{1}{N(1-z)}+\frac{1}{\pi}\log cN -g_A + \cdots
\label{Gg}
\end{equation}
where as before $g_A$ stands for $g_A(1)$  and the dots denote higher
order terms.
As for the infinite lattice, our procedure will be to bring the 
summation on $A$ outside the integration on $z$ and
shift the integration path around the poles of the integrand. Therefore
we should now discuss these poles. All required knowledge about the
behavior of the various quantities involved has been collected in
Sec.\,\ref{zto1}.
First, it follows from Eq.\,(\ref{Gg}) and the definition (\ref{defcn})
of $m_0$ that for $z\rightarrow 1$
\begin{equation}
\sum_{A\neq\emptyset}\mu_A\frac{1}{G_0(z)-g_A(z)} \simeq -Nm_0(1-z)
\label{sumGg}
\end{equation} 
Hence the integrand of Eq.\,(\ref{resultMtaverage}) has a simple pole at
$z=1$ with residue $Nm_0$. 
Secondly, as is clear from the
discussion in Sec.\,\ref{zto1}, this integrand has 
special simple poles for $z=z_A$, where
\begin{equation}
z_A=1+\frac{\pi}{N\log cN}\Big[ 1+\frac{\pi g_A}{\log cN} + 
{\cal O}(\frac{1}{\log^2\! N}) \Big]
\label{zA}
\end{equation}
These are the only poles at a distance of ${\cal
O}(1/N\log N)$ from $z=1$, all the other ones being at least at
distances of
${\cal O}(1/N)$. Therefore, on time scales that are  
at least of ${\cal O}(N\log N)$, the other poles will contribute
vanishingly to the result.

Carrying the integral out but retaining only the contribution of the poles at
$z=1$ and $z=z_A$ leads to
\begin{eqnarray}
\overline{M}(t)&\simeq&Nm_0 \nonumber\\
& &+N\sum_{A\neq\emptyset}\mu_A \exp\Big[-\frac{\pi t}{N\log cN}
\Big(1+\frac{\pi g_A}{\log cN}+ {\cal O}(\frac{1}{\log^2\! N}) \Big)
\Big] \nonumber\\
& &
\label{Mtworegimes}
\end{eqnarray}
The first term in this equation is not present for the black-and-white
observables, which have $m_0=0$. We shall discuss these observables first.
Two different time scales are of interest.\\

\noindent {\it 1. The main regime,}\, in which an observable takes values
of the same order as its maximum value.
To focus on this regime we scale time as 
\begin{equation}
\tau=\frac{\pi t}{N\log cN}
\label{deftau}
\end{equation}
and take the limit $N\!\rightarrow\!\infty, \; t\!\rightarrow\!\infty$
at $\tau>0$ fixed. In this "{\it $\tau$-limit}\,"
Eq.\,(\ref{Mtworegimes}) 
leads directly to   
\begin{equation}
\overline{M}(t)\simeq \frac{\pi N}{\log cN}\, m_1\, \tau\,
\text{e}^{-\tau} 
\label{Mmainregime}
\end{equation}
with $m_1$ defined by Eq.\,(\ref{defcn}).\\

\noindent {\it 2. The long-time regime,}\, in which the observable
approaches its final value. It is now appropriate to scale time as
\begin{equation}
\sigma=\frac{\pi t}{N\log^2\! cN}
\label{defsigma}
\end{equation}
with $\sigma>0$. In this "{\it $\sigma$-limit}\,"
Eq.\,(\ref{Mtworegimes}) gives
\begin{equation}
\overline{M}(t)\simeq N(cN)^{-\sigma}\sum_{A\neq\emptyset}\mu_A\text{e}^{-\pi g_A
\sigma} 
\label{Mlongtime}
\end{equation}
One may notice that the result (\ref{Mmainregime}) is recovered from
Eq.\,(\ref{Mlongtime}) if one sets $\sigma=\tau/\log cN$ and expands the
terms inside the sum on $A$ to first order in $1/\log cN$.

For black observables Eq.\,(\ref{Mtworegimes}) with the scaling of
Eq.\,(\ref{deftau}) leads directly to 
\begin{equation}
\overline{M}(t)\simeq Nm_0(1-\text{e}^{-\tau})
\label{Mblack}
\end{equation}
This decay law depends on the type of the black pattern only
through the prefactor $m_0$. 

\subsection{Examples}
\label{examples2}
We consider the same examples as in Sec.\,\ref{examples} but now on the
finite lattice.\\

\noindent{\it Example 2.1.}\, Let $M=S$. Substituting $m_0=1$ in
Eq.\,(\ref{Mtworegimes}) we see that on a finite lattice of $N$ sites
the average number of unvisited sites decays as
\begin{equation}
N-\overline{S}(t)\simeq N\text{e}^{-\tau}
\label{N-S}
\end{equation}
a result first obtained by Weiss {\it et al.}
\cite{WeissHavlinBunde}. It is also valid in the
long-time regime, where it
can be written as $N(cN)^{-\sigma}$. In this regime, since $\sigma>0$, the
unvisited sites 
constitute an infinitesimally small fraction of all lattice sites.\\

\noindent{\it Example 2.2.}\, Let now $M=E$. Using in
Eq.\,(\ref{Mtworegimes}) the appropriate coefficients $\mu_A$ from 
Table I we obtain
\begin{align}
\overline{E}(t)\simeq &\;\frac{2\pi N}{\log
cN}\,\tau\,\text{e}^{-\tau}\nonumber\\
\overline{E}(t)\simeq &\;4 N(cN)^{-\sigma}(1-\text{e}^{-\pi\sigma/2})
\label{Emainlong}
\end{align}
in the main and long-time regimes, respectively. The prefactor
$N(cN)^{-\sigma}$ in the second one of these equations is the average
number of unvisited sites found in the preceding
example. If these sites were randomly distributed, then since they are
infinitely dilute, each of them would have four visited neighbors and
the result for $\overline{E}(t)$ would be only $4N(cN)^{-\sigma}$. Hence
the factor $1-$e$^{-\pi\sigma/2}$ 
represents nontrivial correlations due to unvisited sites clustering
together.\\ 

\noindent{\it Example 2.3.}\, Let $M=I$. Using in
Eq.\,(\ref{Mtworegimes})
the coefficients $\mu_A$ from Table I appropriate to this case, we find
\begin{align}
\overline{I}(t)\simeq&\:\frac{(\pi/2-1) N}{\log cN}\tau e^{-\tau}\nonumber\\
\overline{I}(t)\simeq&\:N(cN)^{-\sigma}(1-2\text{e}^{-\pi\sigma/2}
+\text{e}^{-(\pi+2)\sigma/2})
\label{Imainlong}
\end{align}
for the main regime and the long-time regime, respectively. Again, the
factor in parentheses in the last equation is due to correlations in the
positions of the unvisited sites.

This example requires the following remark.
The expression of Eq.\,(\ref{exprI}) for the number of islands $I(t)$ is
correct only for the infinite lattice. For the finite lattice with
periodic boundary conditions in both directions one should add $-1$ 
when the support closes onto itself around the torus in one of the
directions and $-2$ 
when it does so in both directions.
In the latter case the first term in Eq.\,(\ref{exprI}) is absent and one
obtains the expression that we used to derive Eqs.\,(\ref{Itfinal}) and
(\ref{Imainlong}). The extra term $-2$ is of course of no importance in
the main 
regime, but it needs to be taken into account in the long-time regime 
to ensure that
$\overline{I}(t)$ vanishes when $t$ tends to infinity.

If we neglect the difference between averages of
ratios and ratios of averages, then we have from Eqs.\,(\ref{N-S}),
(\ref{Emainlong}), and (\ref{Imainlong}) that the quantities
\begin{align}
[N-\overline{S}(t)]/\overline{I}(t)\simeq &\:1/(1-2\text{e}^{-\pi\sigma/2}
+\text{e}^{-(\pi+2)\sigma/2})\nonumber\\
\overline{E}(t)/\overline{I}(t)\simeq &\:4(1-\text{e}^{-\pi\sigma/2})/
(1-2\text{e}^{-\pi\sigma/2}
+\text{e}^{-(\pi+2)\sigma/2})
\label{ratios}
\end{align}
represent the average area and the average perimeter, respectively, of
an island. In the main time regime the expression for the average area 
simplifies to
\begin{equation}
[N-\overline{S}(t)]/\overline{I}(t)\simeq \frac{2\log
cN}{\pi-2}\:\tau^{-1} 
\label{tau-1}
\end{equation}
It is now possible to make a comparison with the simulations by Coutinho
{\it et al.} \cite{Coutinho}, carried out on finite square lattices
of up to $N=1200^2$ sites with periodic boundary conditions. These
authors were interested in the "fragmentation" of the finite lattice 
into islands and the way the average number and size of the islands 
eventually tend to zero.
The comparison leads to the following conclusions.

({\it i})\, In an {\it early time regime}, which for a 
lattice of $600^2$ sites
corresponds to $t$ less than $\,\approx
0.5\times 10^6$, the finite lattice size still
plays no role and our Eq.\,(\ref{Itfinal}) for $\overline{I}(t)$ agrees
within error bars with the simulation data shown in 
Ref.\,\cite{Coutinho} for 
$N=600^2$. These data are for $t>0.05\times 10^6$; for the agreement
to be of this quality, not only the leading order term
but also the two correction terms in Eq.\,(\ref{Itfinal}) have to be 
taken into account. 

({\it ii})\, In the early time regime and in the main regime, where 
$\overline{I}(t)$ passes
through its maximum, our long time expansion Eq.\,(\ref{Imainlong}) 
overestimates the numerical values of $\overline{I}(t)$ by up to 50\%; 
in the long 
time regime, when $\overline{I}(t)$ starts to decay, the agreement with 
the simulation data becomes rapidly better and stays good all the way up to 
$t\approx 19\times
10^6$, where with a large probability no unvisited sites are left. 

({\it iii})\,  Coutinho {\it et al.} in their simulation find the time 
dependent average island size on a 
lattice of $N$ sites to be a function only of the scaling variable 
$\sigma$. Our expression (\ref{ratios}) confirms this result.
The scaling function is not, however, the power law that it was thought
to be in
Ref.\,\cite{Coutinho}, but the inverse of a sum of exponentials given in
Eq.\,(\ref{ratios}). According to our Eq.\,(\ref{tau-1}) a power law 
appears only in the main regime and has the exponent $-1$.
In the long time regime Eq.\,(\ref{ratios}) leads to an apparent
exponent with larger absolute value.

({\it iv})\, In the preceding discussion we have compared the
numerical data for the average island size $\overline{[N-S(t)]/I(t)}$
to the theoretical result (\ref{ratios}) for
$[N-\overline{S}(t)]/\overline{I}(t)$ that we were able to calculate. 
We have not been able to calculate directly the average island size. Nor
have we been able to calculate still another quantity determined in the
simulation of Ref.\,\cite{Coutinho}, {\it viz.} the time dependent
"diversity" of the  
island sizes, defined as the number of different sizes that occur at any
given time.


\section{Fluctuations and correlations}
\label{fluctuations}
\subsection{Relation to a first passage time problem}
\label{relfirst2}
The original determination by Dvoretzky and
Erd\"os \cite{DvoretzkyErdos} of the average 
number $\overline{S}(t)$ of lattice sites in the support of a random
walk was followed 
only much later \cite{JainPruitt,Torney} 
by a calculation of the root-mean-square deviation of this quantity from
its average.
Yet that calculation was important, because the result, exhibited in our
Eq.\,(\ref{Svariance}), shows that in the limit
$t\rightarrow\infty$ the probability distribution of $S(t)$
becomes infinitely narrow, even though only
logarithmically slowly with $t$.

It is now natural to ask 
if the more general observables whose averages we studied in
Sec.\,\ref{calcaverages} also have infinitely narrow distributions for
$t\rightarrow\infty$. Without much extra effort
it will be possible to calculate also the cross-correlations. 
We therefore consider
two observables of the form (\ref{Generalform}), namely
\begin{equation}
M(t)=\sum_A\mu_A M_A(t),\qquad M'(t)=\sum_B\mu^\prime_BM_B(t)
\label{MandM'}
\end{equation}
which have $\sum_A\mu_A=\sum_B\mu^\prime_B=0$, and focus on
\begin{equation}
\overline{\Delta M(t)\Delta M'(t)}\, \equiv\, \overline{M(t)M'(t)}
-\overline{M}(t)\;\overline{M'}(t)
\label{DMDM'av}
\end{equation}
In this section we confine our analysis to the
infinite lattice.\\

The first step will be to find an expression for the generating function 
$\hat{C}_{MM'}$ defined by
\begin{equation}\label{GenFctFluct}
\hat{C}_{MM'}(z)=\sum_{t=0}^\infty z^t\,\overline{M(t)M'(t)}
\label{defCz}
\end{equation}
When working out the RHS of Eq.\,(\ref{GenFctFluct}) with the aid of
Eqs.\,(\ref{MandM'}) and (\ref{MA(t)}) we encounter averages of products
$m_{\x+A}(t)m_{\y+B}(t)$ where $\x$ and $\y$ are arbitrary
lattice vectors. It is then convenient to write $\y=\x+\r$ and to
define $U(\r)$ as the union of $A$ and $\r+B$. 

$A$ and $B$ are finite sets, small in most applications, and we may
locate them near the origin without loss of generality. For
$r$ not too small, therefore, $U(\r)$ will be the union
of {\it disjoint} sets $A$ and $\r+B$, and its number of elements
will be $|U|=|A|+|B|$. 

With this notation  we get
\begin{equation}
\overline{M(t)M'(t)}=\sum_{A,B}\mu_A\mu_B^\prime \sum_{\r}
\overline{M}_{U(\r)}(t)  
\label{CMM't}
\end{equation}  
The observable $M_{U(\r)}$ that occurs here is exactly of the form
(\ref{definMA}), with $A$ replaced by $U(\r)$.\\  

The generating function $\hat{C}_{MM'}(z)$ is now easily expressed in
terms of the first passage probabilities defined in
Sec.\,\ref{otherobservables}. The derivation is analogous to the one of
Eq.\,(\ref{exprMbarz}) and the result is
\begin{eqnarray}
\hat{C}_{MM'}(z)&=&-\frac{1}{1-z}\sum_{A,B}{}^\ast\mu_A\mu_B^\prime\sum_{\r}
\sum_{\bu\in U(\r)}\hat{F}_{U(\r)}(\bu,z)\nonumber\\
&=&-\frac{1}{1-z}\sum_{A,B\neq\emptyset}\mu_A\mu'_B\sum_{\r}\Big[\sum_{\bu\in
U(\r)}\hat{F}_{U(\r)}(\bu,z)\nonumber\\
& &\phantom{-\frac{1}{1-z}} - \sum_{\ba\in A}\hat{F}_A(\ba,z)
-\sum_{\bb\in B}\hat{F}_B(\bb,z)\Big]
\label{CMM'z}
\end{eqnarray} 
in which the asterisk indicates that the term with $A=B=\emptyset$ is
excluded from the summation.
The function $\hat{F}_{U(\r)}(\bu,z)$ in Eq.\,(\ref{CMM'z}) should be
determined from the linear system of equations
\begin{equation}
\sum_{\bu'\in
U(\r)}\hat{F}_{U(\r)}(\bu',z)\hat{G}(\bu-\bu',z)=\frac{1}{1-z}
\label{eqnforFU}
\end{equation}
for all $\bu\in U(\r)$. Formally this equation is strictly analogous to
Eq.\,(\ref{eqnforF}), 
but the structure of the matrix involved is different. It depends on
the parameter $\r$, which represents the distance between the two
components of the set $U(\r)$.

\subsection{Scaling limit and long time behavior}
\label{sclimitlongtime}
We shall be able to solve
$\hat{F}_{U(\r)}$ from Eq.\,(\ref{eqnforFU}) 
only in the scaling limit $z\rightarrow 1$,
$r\rightarrow\infty$, with $\xi^2\equiv 4 r^2(1-z)$ fixed. But since
$z\rightarrow 1$ is exactly the limit of interest, this solution
suffices provided the sum on $\r$ in Eq.\,(\ref{CMM't}) is dominated by
values $r\sim (1-z)^{-1/2}$. In the scaling limit we need to consider
only the case where $A$ and $\r+B$ are
disjoint. Using the shorthand notation of
Eq.\,(\ref{matrixshorthand}) we find from Eq.\,(\ref{eqnforFU}) 
\begin{eqnarray}
\sum_{\ba'\in A}(1-\gamma_{\ba\ba'}(z)){\cal
F}_{\ba'}+\sum_{\bb'\in B}\gamma_{\ba,\r+\bb'}(z){\cal
F}_{\r+\bb'}&=&1\nonumber\\
\sum_{\ba'\in A}\gamma_{\r+\bb,\ba'}(z){\cal
F}_{\ba'}+\sum_{\bb'\in B}(1-\gamma_{\bb\bb'}(z)){\cal
F}_{\r+\bb'}&=&1
\label{matrixeqnforcalFU}
\end{eqnarray}
valid for all $\ba\in A$ and $\bb\in B$, with $\cal F$ now defined on
the set $U(\r)$. In the scaling limit
Eqs.\,(\ref{Gscaling}) 
and (\ref{matrixshorthand}) imply the behavior
\begin{equation}
\gamma_{\ba,\r+\bb}(z)\simeq 2\pi^{-1}K_0(2r\sqrt{1-z})/G_0(z)+{\cal O}(1-z)
\label{scgamma}
\end{equation}
whose important feature  is that to leading order no dependence on
$\ba$ or $\bb$ appears. Furthermore
\begin{eqnarray}
\gamma_{\ba\ba'}(z)\simeq\frac{g(\ba-\ba')}{G_0(z)},\qquad\gamma_{\bb\bb'}(z)
\simeq\frac{g(\bb-\bb')}{G_0(z)}
\label{gammaz=1}
\end{eqnarray}
and we shall henceforth suppress the argument of the $\gamma$'s. 

In the  scaling limit Eqs.\,(\ref{matrixeqnforcalFU}) can then be 
written, in matrix notation, as 
\begin{equation}
\left(
\begin{array}{cc}
J^{[\alpha,\alpha]}-\gamma^{(A)}&\lambda J^{[\alpha,\beta]}\\\lambda
J^{[\beta,\alpha]}&J^{[\beta,\beta]}-\gamma^{(B)}
\end{array}\right){\cal F}=\text{\bf j}
\label{sceqnforcalFU}
\end{equation}
where $\alpha=|A|$,\; $\beta=|B|$,\;
$\gamma^{(A)}$ and $\gamma^{(B)}$ are matrices as defined in 
Sec.\,\ref{firstpassage},\; $J^{[\ell,m]}$ is the $\ell\times m$ matrix
with all elements equal to 1, $\text{\bf j}$ is the vector with all
elements equal to 1, and
\begin{equation}
\lambda(\xi,z)=2\pi^{-1}K_0(\xi)/G_0(z)
\label{deflambda}
\end{equation} 
We can obtain from Eq.\,(\ref{sceqnforcalFU}) a formal expression, 
analogous to 
Eq.\,(\ref{solnsumcalF}), for the sum of the components of $\cal F$. In
the appendix it is shown that this expression can be transformed to
\begin{equation}
\sum_{\bu\in U(\r)}\!{\cal F}_{\bu}=\frac{2-2\lambda
-\gamma_{A}-\gamma_B}{1-\lambda^2-\gamma_A-\gamma_B+\gamma_A\gamma_B}
\label{solncalFU}
\end{equation}
Upon substituting Eq.\,(\ref{solncalFU}) in 
Eq.\,(\ref{CMM'z}) and replacing the
sum on $\r$ that occurs there by an integral on $\xi$ we find
\begin{equation}
\hat{C}_{MM'}(z)\simeq
-\frac{\pi}{2(1-z)^3G_0(z)}\sum_{A,B\neq\emptyset}\mu_A\mu_B^\prime
\int_0^\infty\!\!d\xi\;\xi\; I(\xi,z)
\label{CMMz}
\end{equation}
in which the function $I(\xi,z)$ is given by
\begin{equation}
I=\frac{2-2\lambda
-\gamma_A-\gamma_B}{1-\lambda^2-\gamma_A-\gamma_B+\gamma_A\gamma_B}
-\frac{1}{1-\gamma_A}-\frac{1}{1-\gamma_B}
\label{integralI}
\end{equation}
It is useful to observe that the three quantities $\gamma_A$,
$\gamma_B$, and $\lambda(\xi,z)$ are all of order $G^{-1}_0(z)$.\\
The steps that follow are again analogous to the procedure of
Sec.\,\ref{longinfinite}. We wish to 
expand $I(\xi,z)$ in inverse powers of 
$G_0(z)$ and substitute the result in Eq.\,(\ref{CMMz}).
The $\mu_B'$ that characterize the observable $M'$ define
coefficients $m'_n$ analogous to the $m_n$ of Eq.\,(\ref{defcn}).
In the expansion we encounter furthermore the coefficients 
\begin{equation}
a_n=\int_0^\infty d\xi\;\xi\;K_0^n(\xi)
\label{defan}
\end{equation}
of which we shall need the explicit values $a_1=1$ and $a_2=\frac{1}{2}$,
as well as \cite{JainPruitt,Torney}
\begin{equation}
a_3=-\frac{1}{2}\int_0^1 d\xi\;\frac{\log\xi}{1-\xi+\xi^2}=0.58597...
\label{valuea3}
\end{equation}
Finally, before working out this expansion it is useful to classify the
observables $M$ 
according to their order. We shall say that $M$ is of {\it order} $k$ if
\begin{equation}
m_0=m_1=\cdots=m_{k-1}=0 \quad\mbox{ and }\quad m_k\neq 0
\label{deforderk}
\end{equation}
Observables of order $k=0$ and $k=1$ have occurred in the preceding sections,
and physically interesting examples with $k\geq 2$ perhaps exist.  
Let now $k$ and $k'$ be the orders of $M$ and $M'$, respectively.
We anticipate -- as will be confirmed by the calculation -- that we have
to expand $\hat{C}_{MM'}(z)$ in Eq.\,(\ref{CMMz}) to order
$1/G_0^{k+k'+4}(z)$. In view of Eqs.\,(\ref{deforderk}) and (\ref{defcn})
only those terms 
in the expansion will survive the summation on $A$ and $B$ that contain
at least a factor $\gamma_A^k$ and a factor $\gamma_B^{k'}$. Since there is one
factor $1/G_0(z)$ outside the sum in Eq.\,(\ref{CMMz}), this leaves room
for at most three factors $\lambda$. We have therefore found it convenient 
to begin by expanding
$I=I'\lambda+I''\lambda^2+I'''\lambda^3+\cdots$ and then
to determine the first three coefficients of this series in terms of
$\gamma_A$ and $\gamma_B$. 
Then the $1/G_0(z)$ expansion of $\hat{C}_{MM'}(z)$ leads to
\begin{align}
\hat{C}_{MM'}(z)\simeq&\frac{1}{(1-z)^3G_0^{k+k'+2}(z)}\Big[
2a_1m_km_{k'}^\prime\nonumber\\
-&G_0^{-1}(z)\Big(\pi^{-1}(k+k'+2)a_2m_km_{k'}^{\prime}-
2a_1(m_km^\prime_{k'+1}+m_{k+1}m_{k'}^\prime)\Big)\nonumber\\
+&G_0^{-2}(z)\Big(8\pi^{-2}(k+1)(k'+1)a_3 m_k m_{k'}^\prime\nonumber\\
&\phantom{G_0^{-2}(z)\Big(}-2\pi^{-1}(k+k'+3)a_2(m_km_{k'+1}^\prime+
m_{k+1}m_{k'}^\prime)\nonumber\\  
&\phantom{G_0^{-2}(z)\Big(}+2a_1(m_km_{k'+2}^\prime+m_{k+1}m_{k'+1}^\prime+
m_{k+2}m_{k'}^\prime)\Big)\nonumber\\
&+\cdots\Big]
\label{CMMZfinal}
\end{align} 
The next step is to invert Eq.\,(\ref{defCz}). After integrating on $z$ with 
the aid of Eq.\,(\ref{asptJnl}) one finds
\begin{eqnarray}
& &\overline{M(t)M'(t)} \simeq \,\frac{\pi^{k+k'+2} t^2}{\log^{k+k'+2}
8t}\Big[a_1m_km_{k'}^\prime\nonumber\\
& &+\frac{1}{\log
8t}\Big(\!\!-\!(k+k'+2)(2a_1b_{13}+a_2)m_km_{k'}^\prime+\pi
a_1(m_km_{k'+1}^\prime+m_{k+1}m_{k'}^\prime)\!\Big)\nonumber\\
& &+\frac{1}{\log^2 8t}
\Big(\{(k+k'+2)(k+k'+3)(a_1b_{23}+2a_2b_{13})\nonumber\\
& &\phantom{\frac{1}{\log^2 8t}\Big(\{} 
+4(k+1)(k'+1)a_3)\}m_km_{k'}^\prime\nonumber\\
& &\phantom{\frac{1}{\log^2 8t}\Big(}
-(k+k'+3)\pi(a_2+2a_1b_{13}) 
(m_km_{k'+1}^\prime+m_{k+1}m_{k'}^\prime)\nonumber\\
& &\phantom{\frac{1}{\log^2 8t}\Big(}
+\pi^2 a_1(m_km_{k'+2}^\prime+m_{k+1}m_{k'+1}^\prime+
m_{k+2}m_{k'}^\prime)\Big)\nonumber\\
& &+\cdots\Big]
\label{MMtfinal}
\end{eqnarray}
From Eq.\,(\ref{Mtfinal}) we deduce that when $M$ is of order $k$ its
average is given by
\begin{align}
\overline{M}(t)\simeq&\frac{\pi^{k+1}t}{\log^{k+1}8t}\Big[m_k+ 
\frac{1}{\log 8t}\Big(-(k+1)b_{12}m_k+\pi m_{k+1}\Big)\nonumber\\ 
&+\frac{1}{\log^2 8t}\Big(\tfrac{1}{2}(k+1)(k+2)b_{22}m_k
-\pi(k+2)b_{12}m_{k+1}+\pi^2m_{k+2}\Big)\nonumber\\
&+\cdots\Big]
\label{Morderk}
\end{align}
Eq.\,(\ref{Morderk}), its counterpart for $\overline{M'}(t)$, and
Eq.\,(\ref{MMtfinal}) now have to be combined in Eq.\,(\ref{DMDM'av}).
Using the
explicit expressions for the coefficients $a_1, a_2, b_{12}, b_{13}$, 
$b_{22}$, and $b_{23}$ leads to the desired 
correlation function. 
For $t\rightarrow\infty$  the two
leading orders cancel in the subtraction in Eq.\,(\ref{DMDM'av}). 
The final result is
\begin{equation}\label{ConnCorr}
\frac{1}{(k+1)(k'+1)}
\frac{\overline{\Delta M(t)\Delta M'(t)}}{\overline{M}(t)\overline{M'}(t)}
\simeq \frac{\cal{A}^2}{\log^2 8t} + \cdots
\label{MMomega}
\end{equation}
where the dots indicate terms of higher order in $1/\log 8t$ and 
\begin{equation}
\cal{A}^2=4a_3+1-\tfrac{1}{6}\pi^2=1.69897... 
\label{valueunivcst}
\end{equation}
Eq.\,(\ref{ConnCorr}) contains as a special case the
well-known result of 
Eq.\,(\ref{Svariance}) for the variance of $S(t)$, originally 
due to Jain and Pruitt \cite{JainPruitt}, and rederived with the aid of
a method more similar to ours by Torney \cite{Torney}.

\subsection{Conclusions}
\label{conclusions}
It is remarkable that the ratio
on the RHS of Eq.\,(\ref{ConnCorr}) is universal: It is independent of
the choice of the 
observables $M$ and $M'$. But we shall now see that
Eq.\,(\ref{ConnCorr})   
has consequences that reach far beyond this simple fact.

For $k=0,1,2,\ldots$ we define for each observable $M$ of order $k$ 
its normalized deviation from average, $\eta_M$, by
\begin{equation}
\eta_M(t) = \frac{\log 8t}{(k+1)\cal{A}}\frac{\Delta M(t)}{\overline{M}(t)}
\label{defetaM}
\end{equation}
These variables satisfy to leading order
\begin{equation}
\overline{\eta_M(t)}=0,\quad\overline{\eta_M(t)\eta_{M'}(t)}=1 \qquad
\mbox{ for all } M,\,M'   
\label{corretaM}
\end{equation}
It follows that for any two $M$ and $M'$ the difference
$\eta_M(t)-\eta_{M'}(t)$ has zero variance, and therefore the
$\eta_M(t)$ {\it are all equal to a single random variable} that we shall
call $\eta(t)$.
As a consequence we can relate the deviation from average $\Delta M(t)$ of any
observable $M(t)$ of order $k$ to $\eta(t)$ by 
\begin{equation}
\Delta M(t)\simeq(k+1)\frac{\cal{A}}{\log 8t}\,\overline{M}(t)\,\eta(t)
\label{DMeta}
\end{equation}  
Upon writing down this equation for the special case $M=S$, using the
explicit expression (\ref{Stfinal}) for $\overline{S}$, and eliminating
$\eta(t)$, one finds
\begin{equation}
\Delta M(t)\simeq(k+1)\frac{\pi^k}{\log^k 8t}\,m_k\Delta S(t)
\label{DMDS}
\end{equation}
This last equation embodies one of the main conclusions of this work:
{\sl All the observables $M$ fluctuate 
around their averages in strict proportionality with the
fluctuation of total number of sites $S(t)$ in the support.}
This conclusion applies, in particular, to the pattern numbers
$N_\alpha(t)$, the total perimeter length $E(t)$ of the support, the
total number $I(t)$ of
islands enclosed by it, and the total number $I_\beta(t)$
of islands of a specific type $\beta$. We are not aware of any computer
simulations that confirm Eq.\,(\ref{DMDS}), although they would be easy to
carry out. 

There is a still different and instructive way to formulate this 
conclusion. Let $M$ be an observable of order $k$ and $\rho$ a quantity
that remains of $\cal{O}(1)$ when $t\rightarrow\infty$.
Eq.\,(\ref{Morderk}) can now be used to establish the asymptotic expansion in
powers of $1/\log 8t$ of $\rho^{-1}\overline{M}(\rho t)$.
Upon comparison with Eq.\,(\ref{DMeta}) and choosing
$\log\rho=-\cal{A}\eta(t)$ one finds that all
fluctuating observables $M(t)$ can be written to second order in the form
\begin{equation}
M(t)\simeq\text{e}^{\cal{A}\eta(t)}\overline{M}(\text{e}^{-\cal{A}\eta(t)}t)
\label{renormalizedtime}
\end{equation}
In the mathematical
literature on Brownian motion (for a review see Le Gall
\cite{LeGallcourse}) the quantity 
$-\cal{A}\eta/(2\pi)$ has appeared in the study of the asymptotic behavior
of the volume of the Wiener sausage (where it is commonly denoted by the
symbol $\gamma$) and is known as the {\it renormalized
local time of 
self-intersections}, a concept introduced by Varadhan in
an appendix to an article by Symanzik \cite{Varadhan}.

\section{Discussion}
\label{discussion}

In a preceding letter \cite{CaserHilhorst} a more general approach was
presented to the much more restricted problem
of how to calculate the average $\overline{I}(t)$. The
total number of islands was written as $I(t)=C(t)-D(t)$, where the
increments $\Delta C(t)=C(t)-C(t-1)$ and $\Delta D(t)=D(t)-D(t-1)$ are
the numbers of islands created and destroyed, respectively, in the $t\,$th
step (either $\Delta C(t)$ or $\Delta D(t)$ vanishes). Hence $C(t)$ and
$D(t)$ are the total number of islands created and destroyed,
respectively, up until time $t$. It was shown \cite{CaserHilhorst} that 
asymptotically for $t\rightarrow\infty$ to leading order
\begin{equation}
\overline{C}(t)\simeq\overline{D}(t) \simeq A\frac{\pi t}{\log 8t}
\label{CDaspt}
\end{equation}
where $A=0.1017...$ In
the difference $\overline{I}(t)=\overline{C}(t)-\overline{D}(t)$ the
leading order (\ref{CDaspt}) cancels and the result (\ref{Itfinal})
appears. It does not seem possible to express $C(t)$ and $D(t)$ as
observables of the type $M(t)$.
The additional determination of $\overline{C}(t)$ and $\overline{D}(t)$
makes the calculation of Ref.\,\cite{CaserHilhorst} more involved. 
Nevertheless, the generating function found there
(Eq.\,(5) of Ref.\,\cite{CaserHilhorst}) for 
$d^2\overline{I}(t)/dt^2$ is equivalent 
to the one for
$\overline{I}(t)$ implied by Eq.\,(\ref{resultMtaverage}) of this work.\\

The success of the study presented here is due to the generating
function method, whose potential is fully exploited. We also run into
what may be the limitations of this method.
There are several quantities that appear naturally but whose averages
and variances we do not
see a way to determine. These include the total external perimeter
discussed in Sec.\,\ref{examples} and the area of the
islands enclosed by the support. 

Many of the quantities discussed in the preceding chapters have close
analogs in planar Brownian motion. We shall denote these analogs by the
superscript $B$. The Brownian motion analog of the support of
the lattice random walk is the set of points ${\cal S}_b^B(t)\subset
\Bbb{R}^2$ that has been swept out in the time interval $[0,t]$ by a disk of
radius $b$ performing Brownian motion with diffusion constant $D$. The
set ${\cal S}_b^B(t)$ is commonly called the {\it Wiener sausage}
associated with the Brownian motion trajectory. The total area
$S_b^B(t)$ of this set is analogous to the number of sites $S(t)$ in the
support of the lattice random walk.

The diffusion constant $D$ is defined so that the mean square
displacement equals $4Dt$. It can be scaled away, but we shall keep
it here to facilitate comparisons between results from different
sources. In the mathematical literature one customarily sets
$D=\frac{1}{2}$, whereas the long-time, large-distance limit of the
random walk of this work yields $D=\frac{1}{4}$.

The asymptotic behavior of $\overline{S_b^B}(t)$
was first determined by Leontovitsh and Kolmogorov
\cite{LeontovitshKolmogorov}. Berezhkovskii {\it et al.} give the
complete asymptotic expansion
\begin{equation}
\overline{S_b^B(t)}\simeq\frac{4\pi Dt}{\log\kappa Dtb^{-2}}
\sum_{m=0}^\infty \frac{(-1)^m b_{m2}}{\log^m\kappa Dtb^{-2}}
\label{SBerezhkovskii}
\end{equation}
where $\kappa\equiv 4\text{e}^{-2C}$ and the $b_{m2}$ are as defined by
Eq.\,(\ref{defbml}).
The difference $\Delta S_b^B(t)$ was shown \cite{LeGallcourse} to be a random
variable such that the distribution of $t^{-1}(\log t)^2 \Delta S_b^B(t)$
converges for $t\rightarrow\infty$ to a limit distribution {\it
identical} to the one of $t^{-1}(\log t)^2\Delta S(t)$.

Two remarks can be made about the relations obtained 
by differentiating Eq.\,(\ref{SBerezhkovskii}) with respect to $b$.
First, let $E_b^B(t)$ be the total boundary length of the Wiener sausage
${\cal S}_b^B(t)$. Assuming that $\partial{\cal S}_b^B$ is sufficiently
regular we have
\begin{equation}
E_b^B(t)=\frac{dS_b^B(t)}{db}
\label{ES}
\end{equation}
Upon averaging, using Eq.\,(\ref{SBerezhkovskii}) and setting $b_{02}=1$,
we find from Eq.\,(\ref{ES})
\begin{equation}
\overline{E_b^B}(t)\simeq\frac{8\pi Dtb^{-1}}{\log^2\kappa Dtb^{-2}}
\Big[1+\cdots\Big] 
\label{EBerezhkovskii}
\end{equation}
This expression has the same asymptotic time dependence as
Eq.\,(\ref{Etfinal}) for $\overline{E}(t)$, but the coefficients do not
coincide. 
Secondly, differentiate once more and consider the dimensionless number  
\begin{equation}
J_b^B(t)\defin-\frac{d^2S_b^B(t)}{db^2}
\label{defJ}
\end{equation}
Its average behaves in the large $t$ limit as
\begin{equation}
\overline{J_b^B}(t)\simeq\frac{8\pi Dtb^{-2}}{\log^2\kappa Dtb^{-2}}
\label{Jaspt}
\end{equation}
It is easy to show that {\it if} the boundary $\partial\cal{S}_b^B(t)$
is sufficiently regular (having at least a tangent vector in each
point), then the quantity (\ref{defJ}) is equal to $2\pi(I_b^B(t)-2)$, with
$I_b^B(t)$ the number of islands. However, these regularity conditions are
not satisfied here since the boundary has cusp points. Nevertheless,
comparison of Eqs.\,(\ref{Jaspt}) and (\ref{Itfinal}) shows that
$\overline{J_b^B}(t)$ has the same asymptotic time dependence as
$\overline{I}(t)$ (but with a different coefficient).. 

We now discuss the relation of our work to results that have appeared in
the mathematical literature. Throughout the comparison it should be
borne in mind that whereas those results are rigorous, the ones of this
paper have been obtained by the usual methods of
mathematical physics.

Mountford \cite{Mountford} was the first to study the connected
components of the complement of the Brownian path of a point 
({\it i.e.,} the case $b=0$). For all times $t>0$ the number of these
components is infinite due to the presence of many small ones. However,
one can ask for example what the number $C_\varepsilon(t)$ is of
connected components with
an area larger than a prescribed value $\pi\varepsilon^2$. Le Gall,
strengthening the results due to Mountford, has shown that
for almost all Brownian motion trajectories (with $D=\frac{1}{2}$)
in a fixed time interval 
\begin{equation}
\lim_{\varepsilon\rightarrow 0}\frac{\varepsilon^2\log^2 Dt\varepsilon^{-2}}
{4Dt} C_\varepsilon(t)=2
\label{Caspt}
\end{equation} 
where we have used dimensional analysis to restore the variables $D$ and $t$.
If one now assumes that $C_\varepsilon(t)$ is of the same order of
magnitude as the 
{\it total} number $I_\varepsilon^B(t)$ of
islands in the Wiener sausage associated with the same Brownian motion 
trajectory executed by a disk of radius $\varepsilon$, then
Eq.\,(\ref{Caspt}) has an asymptotic time dependence that agrees with
the result of our Eq.\,(\ref{Itfinal}).

Werner \cite{Werner} considers the {\it shape} of the connected
component containing a prescribed point not on the trajectory, and shows
that for $t\rightarrow\infty$ the probability distribution of this
quantity tends to a well-defined limit law.
This result is possible only because of the scale invariance of Brownian
motion and it would be cumbersome to formulate its lattice counterpart,
even in the long-time, large-distance limit. The statement
of our Eq.\,(\ref{Ialphat}) about the number of islands $I_\beta(t)$ of
given type $\beta$, and the associated result about its fluctuation
$\Delta I_\beta(t)$ implied by the discussion at the end of Sec.\, 5,
constitute the point of closest approach between this paper and Werner's.

Finally, we summarize the new results of this paper.
On the one hand, we give the explicit asymptotic behavior as
$t\rightarrow\infty$ for several new
observables associated with the support of the lattice random walk. These
include the total perimeter length, the total number of islands, and the
total number of single-site and dimer islands. On
an infinite lattice the asymptotic behaviors all consist of a
leading term multiplied by a series in inverse powers of $1/\log8t$.
The perimeter and the number of
islands are also considered on a finite lattice, where scaling laws are
obtained in terms of the time $t$ and the lattice size $N$, and a
comparison with computer simulations by Coutinho {\it et al.}
\cite{Coutinho} is possible.  

On the other hand, there is the important general result of
Sec.\,\ref{conclusions}, stating that
the pattern numbers all fluctuate in strict proportionality with one another
and with the total number $S(t)$ of sites in the support.
The fundamental fluctuating variable, called $\eta$ in this work, is the 
renormalized local time of
self-intersections. This result strongly contributes to shape our
picture of the support of the two-dimensional random walk: At any fixed
time $t$, a large class of 
detailed properties is determined by the value of the single random
variable $\eta$. 

One of the open questions that can now be formulated is
connected exactly with this 
random variable, which appears in our work as the time dependent variable
$\eta(t)$. Existing results seem to concern exclusively its stationary
distribution ${\cal P}(\eta)$, which prevails in the limit
$t\rightarrow\infty$. Our investigations point towards the interest of
also studying the 
autocorrelation function $\overline{\eta(t)\eta(t')}$ and possibly other
time dependent properties.
A second question that naturally comes up is: How does the picture in higher
dimensions differ from the one found here in $d=2$? It should certainly
be expected that in sufficiently high dimension the pattern numbers
become independent random variables. The mechanism by which this
independence comes about seems worthy of further elucidation.
We shall leave these and other questions for future work.

\section*{Acknowledgments}

H.J.H. acknowledges a discussion with Professor
M. Coutinho-Filho that laid the germ out of which part of this work grew, and
thanks him for correspondence on the simulation results for finite
lattices. F.v.W. and H.J.H. thank Dr. M. Yor, Dr. W. Werner, and
Professor G. Lawler for discussions helping them to interpret their
results in the light of the mathematical literature. 


\appendix
\section{Appendix}
In this appendix we prove the matrix algebra results we
use to derive averages and correlations.

The following result is used in Sec.\,\ref{firstpassage}.
Let $A$ denote an invertible matrix of dimension
$\ell\times\ell$, with $\ell\geq 2$. In what follows $J^{[m,n]}$ stands
for the $m\times
n$ matrix whose elements are $J^{[m,n]}_{ij}=1$. Define furthermore
\begin{equation}\label{app0}
g_A^{-1}\equiv\sum_{i,j}(A^{-1})_{ij}=\mbox{Tr}(J^{[\ell,\ell]}A^{-1})
\end{equation}
We achieve our aim of finding a simplified form for 
\begin{equation}\label{app1}
\Gamma_A\equiv \sum_{i,j}[(J^{[\ell,\ell]}+A)^{-1}]_{ij}
\end{equation}
by rewriting Eq.\,(\ref{app1}) as
\begin{equation}\label{app2}
\Gamma_A=\mbox{Tr}\Big[J^{[\ell,\ell]}A^{-1}(\mbox{\bf
1}+J^{[\ell,\ell]}A^{-1})^{-1}\Big]\end{equation}
where $\mbox{\bf 1}$ is the unit matrix.\\ 
An intermediate step of the demonstration consists in noting that 
\begin{equation}\label{appu}
J^{[k,l]}CJ^{[\ell,m]}D=g_{C^{-1}}^{-1}J^{[k,m]}D
\end{equation}where $C$ and $D$ are square matrices of dimensions
$\ell\times\ell$ and $m\times m$, respectively. In Eq.\,(\ref{app2}) we
now expand the argument of the trace in powers of
$J^{[\ell,\ell]}A^{-1}$. Iteration of Eq.\,(\ref{appu}) and use of
Eq.\,(\ref{app0}) lead to
\begin{equation}
\mbox{Tr}\Big[(J^{[\ell,\ell]}A^{-1})^n\Big]=g_A^{-n}
\end{equation}
whence
\begin{equation}
\Gamma_A=\frac{1}{1+g_A}
\label{GamA}
\end{equation}
When $\ell=1$ formula (\ref{GamA}) remains valid for all $A\in \Bbb{R}$
(and in particular for $A=0$) if for that case one supplements 
Eq.\,(\ref{app0}) with $g_A=A$.

A generalized version of the above result is needed in 
Sec.\,\ref{fluctuations}. It involves two invertible
matrices $A$ and $B$ of dimensions $\ell\times\ell$ and
$m\times m$, respectively, with $\ell$,\,$m\geq 2$. We now wish to
find a simplified expression for
\begin{equation}\label{app3}
\Gamma_{AB}(\lambda)\equiv\sum_{i,j}\left[\left(\begin{array}{cc}J^{[\ell,
\ell]}+A&\lambda J^{[\ell,m]}\\\lambda
J^{[m,\ell]}&J^{[m,m]}+B\end{array}\right)^{-1}\right]_{ij}
\end{equation}
Upon using the decomposition
\begin{align}
\left(\begin{array}{cc}J^{[\ell,\ell]}+A&\lambda J^{[\ell,m]}\\\lambda
J^{[m,\ell]}&J^{[m,m]}+B\end{array}\right)=&\left(\begin{array}{cc}\mbox{
\bf 1}+J^{[\ell,\ell]}A^{-1}&\lambda J^{[\ell,m]}B^{-1}\\\lambda
J^{[m,\ell]}A^{-1}&\mbox{\bf
1}+J^{[m,m]}B^{-1}\end{array}\right)\!
\left(\begin{array}{cc}\!A&0\\\!0&B\end{array}
\!\right)\nonumber\\&
\end{align}
and writing the double sum in Eq.\,(\ref{app3}) as a trace with the aid
of the matrix $J^{[\ell+m,\ell+m]}$, one finds
\begin{equation}\label{app4}
\Gamma_{AB}(\lambda)=\mbox{Tr}\Big[M(1)(\mbox{\bf 1}+M(\lambda))^{-1}\Big]
\end{equation}
where 
\begin{equation}\label{app4p}
M(\lambda)\equiv\left(\begin{array}{cc}J^{[\ell,\ell]}A^{-1}&\lambda
J^{[\ell,m]}B^{-1}\\\lambda
J^{[m,\ell]}A^{-1}&J^{[m,m]}B^{-1}\end{array}
\right)
\end{equation}

We now expand the argument of the trace in Eq.\,(\ref{app4}) in powers of
$M(\lambda)$. Hence we are left with the calculation of
$\mbox{Tr}(M(1)M(\lambda)^n)$, which is carried out by finding
recursively in $n$ relations between the blocks of
\begin{equation}
M(1)M(\lambda)^n\equiv
U_n\equiv\left(\begin{array}{cc}X_n&Y_n\\Z_n&T_n\end{array}\right)
\end{equation}
Knowing that $U_{n+1}=U_n M(\lambda)$, one has
\begin{align}
X_{n+1}=&X_nJ^{[\ell,\ell]}A^{-1}+\lambda
Y_nJ^{[m,\ell]}A^{-1}\nonumber\\Y_{n+1}=&\lambda X_nJ^{[\ell,m]}B^{-1}+Y_n
J^{[m,m]}B^{-1}
\end{align}
Hence, again with the use of Eq.\,(\ref{appu}), one proves recursively that
$X_n$ and $Y_n$ are of the form
\begin{align}
X_n=&\xi_n J^{[\ell,\ell]}A^{-1}\nonumber\\Y_n=&\eta_n J^{[\ell,m]} B^{-1}
\end{align}
with initial values $\xi_0=\eta_0=1$ and the recursion relations
\begin{eqnarray}\label{app5}
\xi_{n+1}=g_A^{-1}\xi_n+\lambda g_B^{-1} \eta_n\nonumber\\\eta_{n+1}=\lambda
g_A^{-1} \xi_n+g_B^{-1} \eta_n
\label{app5p}\end{eqnarray}
Upon setting $Z_n=\zeta_n J^{[m,\ell]}A^{-1}$ and
$T_n=\tau_nJ^{[m,m]}B^{-1}$ one finds in a similar way
\begin{eqnarray}\label{app6}
\tau_{n+1}=g_B^{-1}\tau_n+\lambda g_A^{-1}
\zeta_n\nonumber\\\zeta_{n+1}=
\lambda g_B^{-1} \tau_n+g_A^{-1} \zeta_n
\label{app6p}\end{eqnarray}
with initial values $\tau_0=\zeta_0=1$.
Hence, using that $\mbox{Tr}\,X_n=g_A^{-1}\xi_n$ and $\mbox{Tr}\,T_n=
g_B^{-1}\tau_n$,
we have from Eqs.\,(\ref{app4}) and (\ref{app4p})
\begin{equation}\label{app7}
\Gamma_{AB}(\lambda)=\sum_{n= 0}^{\infty}(-1)^n\mbox{Tr}\,U_n=\sum_{n=
0}^{\infty}(-1)^n(g_A^{-1}\xi_n+g_B^{-1}\tau_n)
\end{equation}
By summing each of the equations (\ref{app5})--(\ref{app6p}) on all $n$
with weight $(-1)^n$ one obtains two sets of linear equations from
which $\sum_{n=0}^\infty(-1)^n\xi_n$ and $\sum_{n=0}^\infty
(-1)^n\tau_n$ may be solved. Substitution in Eq.\,(\ref{app7}) yields
the final result
\begin{equation}\label{GamAB}
\Gamma_{AB}(\lambda)=\frac{2-2\lambda+g_A+g_B}{1-\lambda^2+g_A
+g_B+g_Ag_B}
\end{equation}
By redoing the calculation with $\ell$ or $m=1$
one finds that this expression for $\Gamma_{AB}$ remains valid for those
special cases.


\newpage
\noindent FIGURE CAPTIONS\\

\noindent {\bf Figure 1.}  A random walker
makes $t$ random steps between the centers of neighboring squares. The 
squares visited 
are colored black and constitute the {\it support} of the
walk. In the example of this figure the support encloses four islands of
unvisited sites. The boundary of the support is represented by a heavy line.\\

\noindent {\bf Figure 2.}  Let the boundary of the support be oriented such
that as one proceeds along it the white sites are on the left and the
black ones on the right. Then in (a) the boundary turns through an angle
$+\frac{1}{2}\pi$ and in (b) through an angle $-\frac{1}{2}\pi$. In (c)
the black squares are (by convention) considered to separate the white
squares from one another and the part of the boundary shown makes two
turns through $+\frac{1}{2}\pi$.


\newpage
\begin {tabular}{||l|r r r r|l||}
\hline
\multicolumn{1}{||c|}{$A$}&\multicolumn{4}{c|}
{$\mu_A$}&\multicolumn{1}{c||}{$g_A$}\\
\cline{2-5}
                        & $S$& $E$& $I$&$I_1$& \\
\hline
$\phantom{\{} \emptyset$&  1 &    &    &    &  -- \\
\{{\bf 0}\}             &$-1$&  4 &  1 &  1 &   0 \\
\{{\bf 0},\,$\be_1$\}   &    &$-4$&$-2$&$-4$& $1/2$\\
\{{\bf 0},\,$\be_1,\be_2,\be_1+\be_2$\}
                        &    &    &  1 &    & $(\pi+2)/2\pi$\\
\{{\bf 0},\,$\be_1,\be_2$\}
                        &    &    &    &  4 & $\pi/2(\pi-1)$\\
\{{\bf 0},\,$\be_1,2\be_1$\}
                        &    &    &    &  2 & $\pi/4$\\
\{{\bf 0},\,$\be_1,-\be_1,\be_2$\}
                        &    &    &    &$-4$&
                                      $\pi(\pi-6)/2(\pi^2-6\pi+4)$\\
\{$\be_1,\be_2,-\be_1,-\be_2$\}
                        &    &    &    &  1 &  1 \\
                                     
\hline
\end{tabular}
\vspace{0.5cm}

\noindent Table I.
{\small
\;The coefficients $\mu_A$ for the four observables $S, E, I$ and $I_1$
defined in the text; entries not shown are zero. 
Symmetry under rotations over $\pi/2$ has been
exploited to reduce the length of the table.
The coefficients in each column add up to zero.
The parameter $g_A$ is given for
all sets $A$ that occur; it is undefined for the
empty set $\emptyset$. $A$ itself is defined only 
up to a translation.

}

\end{document}